# Grid-based Search Technique for Massive Academic Publications

Mohammed Bakri Bashir[1], Muhammad Shafie Abd Latiff[2]
[1,2]Department of Computer Science,
Faculty of Computing, Universiti Teknologi Malaysia UTM,
Johor, Malaysia.
mhmdbakri@gmail.com[1], shafie@utm.my[2]

Shafi'i Muhammad Abdulhamid[3], Cheah Tek Loon[4]
[3,4]Department of Computer Science,
Faculty of Computing, Universiti Teknologi Malaysia UTM,
Johor, Malaysia.
shafii.abdulhamid@futminna.edu.ng[3],

*Abstract*— the numerical size of academic publications that are being published in recent years had grown rapidly. Accessing and searching massive academic publications that are distributed over several locations need large amount of computing resources to increase the system performance. Therefore, many grid-based search techniques were proposed to provide flexible methods for searching the distributed extensive data. This paper proposes search technique that is capable of searching the extensive publications by utilizing grid computing technology. The search technique is implemented as interconnected grid services to offer a mechanism to access different data locations. The experimental result shows that the grid-based search technique has enhanced the performance of the search.

## I. INTRODUCTION

In many scientific research sectors, researchers need to access and search distributed data collections of gigabyte and terabyte scale. Moreover, in several cases data collections must be shared by large communities of users that pool their resources from different site location or a large number of institutions [1]. In addition, sharing distributed academic publications is the vital part of a distributed research community, and an efficient search technique to handle the shared data is crucial to making the distributed information available to researchers. However, searching massively distributed data in large-scale communities are very challenging due to the potential large amounts of data, diverseness, distributed arrangement, and dynamic nature in which the data is growing rapidly [2]. Furthermore, Continuous increase of the data size requires more resources to store the newly produced data. Additionally, the execution time depends on the number of the site involved in the search tasks and the number of query that requires simultaneous processing.

Nonetheless, the aforementioned issues can be addressed by using grid technology, which provides a means to access, manage, control, and store the distributed data [3]. The grid technology provides big organizations and scientific centers a computing power in order to solve complex problems [4]. The grid-based data sources have several features such as decentralization, heterogeneous, and dynamic, in which, searching these data sources are a distributed query [5]. The addition of large-scale feature to the grid-based data source makes the search for these data a complex task [6, 7]. Grid computing can handle the dynamicity of the organizations resources that join or leaves the system at any time. Furthermore, grid computing presents the distribution of the data and resources for the end user as one big supercomputer that contains all the data. The searching process is performed as grid job, which monitors and is distributed over the data sites by the grid scheduler.

The paper describes and explains in details the design and implementation of the Grid-based Academic Publications Search (GAPS) technique. The paper starts with the review of the related search technique by using grid computing. The GAPS components, as well as their functions, are explained in details. The experiments and the evaluation section are conducted to validate the search technique. This paper ends with a conclusion section.

## II. RELATED WORK

The distributed shared data will be beneficial if supported with access and search mechanism. The current sharing systems provide diverse ways to implement the search techniques for distributed data based on the grid infrastructure. Furthermore, the majority of the data sources are not in the form of database management system rather it is in a file form that means the query processing will not be useful to search these files.

The techniques such as the ones proposed and discussed in [8-10] [11] [12] [13] [14] proposed searching systems to search the information in a different format depends on the the data of the system. Furthermore, the majority of the data is not a database management system but it is files (XML, HTML, etc…), that means the query processing will not be useful to search these files. Additionally, the searching techniques use grid computing as a tool to facilitate the search and the harvest of information across the federated locations. Moreover, all those systems use grid computing for indexing and searching so as to speedup query execution and to increase the scalability of the techniques.

However, some of these systems suffer from bottleneck problems that affect the response time and the scalability feature of the searching technique. As a result, the ratio of system failure will increase with the number of user queries. Furthermore, most of the proposed researches do not support the real time search engine instead search indexed data. The choose of grid over cloud is because the cloud is an on-demand resource, while the grid is relatively free.

### III. GRID-ENABLER SEARCH TECHNIQUE

The Grid-Based Academic Publications Search technique (GAPS) is a group of modules that communicate with each other to provide a means to search the distributed academic publications as illustrated in Figure 1. The technique is implemented as modules distributed over grid architecture to provide a mechanism to interaction among the VOs. The GAPS was implemented and integrated as grid services to enable data search to run over grid architecture and to orchestrate the interaction over the grid nodes. The local Search Service module was a Java program installed in each worker node, in a grid architecture and was responsible for performing the search process in the local dataset. Additionally, the other modules are implemented on the head node (broker) of every VO.

#### A. GAPS Components

The research integrates two diverse fields namely academic publications search operations and grid services. On one hand, grid is required to provide a platform to facilitate all the grid services such as data transfer, data location, and data replication. On the other hand, the search operations have been determined after studying different search applications and after identifying the nature of these search techniques as well as the requirement of the grid to work under these techniques. The design and the functionality of the search components will be discussed in the following sections.

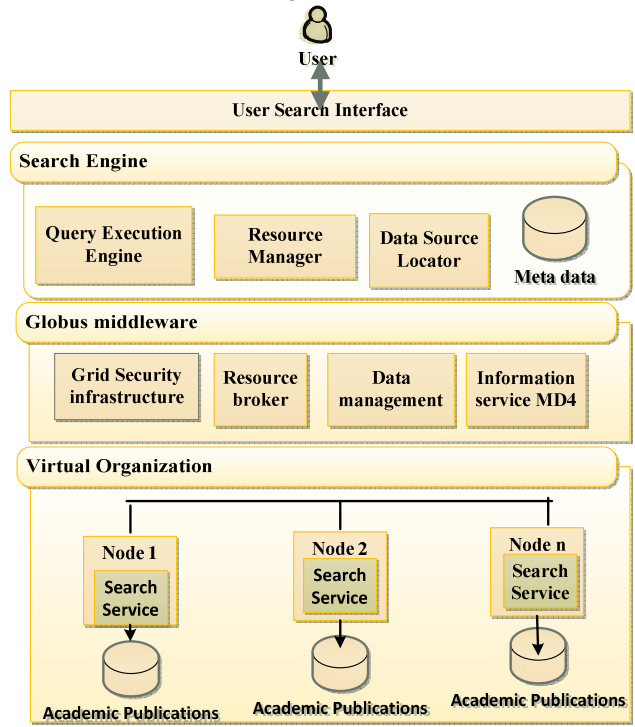

Figure 1: Grid-enabler Search components.

*1) Query Search Engine (QEE)*

QEE is the component that orchestrates and coordinates the query execution over the grid nodes. Additionally, the QEE has several instances distributed among VOs, in which each VO have instance of the QEE. This distribution of the services provides a decentralized search execution, which prevents the system from bottleneck and scalability problems. It means that each VO is equipped with one QEE service, and each node in the VO deploys a copy of the local search service. The QEE determines the nodes that will perform a search at run time by utilizing its internal modules.

After the user submits the search text, the QEE will request the resources information from the Resource Manager, who stores the status and all information about system resources. The lists of the data sources that are involved in the search task are gathered from the Data Source Locator component. The list of available resources and data sources are submitted to the QEE to produce the execution plan of the search jobs. The execution plan that distributes the datasets over the nodes depends on the previous performance and produces the best combination to handle the query. The QM executes the search tasks and returns the result of the search to the end user.

*2) Query Manager (QM)*

QM is the component that involves several functions to execute the user query in grid nodes and returns the best result relevant for the query to the user. One of the QM functions is to receive the list of all available resources in the grid that can be used to perform the query. A list of resources is assigned with the suitable data source that provides a better performance. Additionally, the QM creates the Job Description File (JDF) with all jobs that will be distributed over grid nodes. The JDF contains the location of all data sources and the local search services that will participate on the search process. Additionally, the JDF includes the user query text as well as the location that should receive the result of the search. Furthermore, the QM keeps track of all job execution in the system by keeping the job information in the database. After the search task is completed, the QM sends the information about resource performance to the database to be used in the future search tasks.

*3) Search Service (SS)*

The distributed data sources are difficult to be accessed and to be searched by centralized search application. Instead, a centralized application that enables the local search applications to be run on each node have data source and will collect the result of the search. The GAPS implements the SS that runs on each node that participates in the search tasks. The SS is implemented as a grid service and is installed to be run with the globus container. The globus container is run once the node starts, and it continues to run until the node shuts down. By applying this method, the SS does not need to wait time to load on the memory when the node receives search job request. Additionally, running the SS as a grid service saves the time required to starts the SS every time the search is performed. The SS is designed and developed based on the object-oriented technology, which allows a new component to be added easily. This feature offers the SS the ability to support more data types.

*4) User Search Interface*

The User Search Interface (USI) is an interaction mechanism proposed to provide the end user access point to deal with the system as illustrated in Figure 2. Additionally, USI offers the end user interface to perform a search task by using GAPS. The USI provides keyword-based and multivariate-based search types to execute on the grid nodes. The experiment shows that the USI overhead is very small as compared with the response time. Furthermore, the result shows the ability of the USI to deal with several VOs.

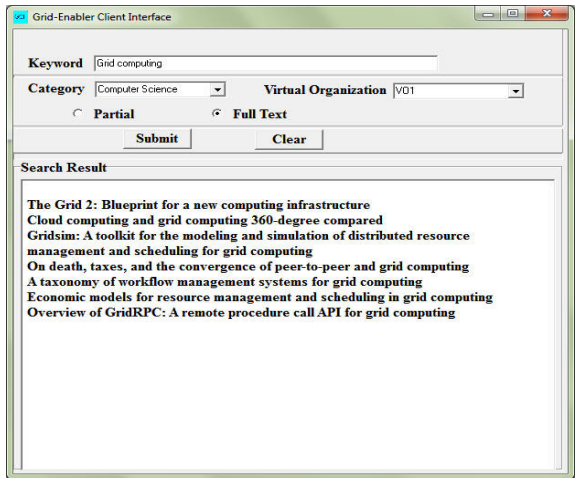

Figure 2: Grid-enabled user search interface.

## IV. EXPERIMENT AND EVALUATION

The goal of this experiment is to investigate the usage of grid computing to search large and distributed academic publications in order to enhance the performance of the search processes. The GAPS is evaluated by using the response time, the speed up, and the efficiency of the search technique.
The experiment conducted in the grid test-bed which contains 12 computer nodes distributed among three Virtual Organizations (VO) and each VO contains four nodes. One of four nodes has two roles as grid broker equipped with Certificate Authority (CA) server and as a computing node. The grid nodes have different specifications by using Red Hat Enterprise Linux 3 as OS. The Globus toolkit 4.0.2 is installed in each broker node because the broker is considered as a CA server.  The datasets used are articles collected from different academic repositories, which contain the open access information about the articles. The worker is equipped with datasets files of different sizes that scale from 10 million records.

*1) Response Time*

This experiment is conducted to measure the search response time by increasing both the data size and the computing nodes. The goal of the experiment is to identify the effect of the GAPS in searching massive academic publications and to measure the enhancement achieved by using the GAPS. Using the GAPS technique as conducted the experiment technique.

Figure 3 shows that the response time starts to decrease when small number of nodes is used and then increases when the number of nodes is more than 5. The result shows that the GAPS has better response time as compared to the traditional search. The response time of the GAPS remains to be faster than the traditional search with 60% while other response time reaches 100%, and some response time decreases to reach 54%. The result shows that the GAPS has fast response time and better performance as compared to the traditional search.

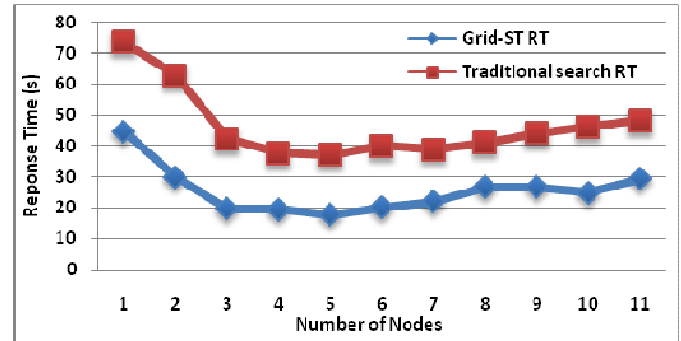

Figure 3: Response time scales as the increase of size.

*2) Speed up*

The speed up is used to measure the performance of the search techniques. The speedup is defined as the ratio of the time to execute the job on a small system until the time to execute the same job on large systems.  An experiment is conducted to measure the speedup of the GAPS when the number of nodes is increased. The speedup is defined as :

$$\text{Speedup} = \frac{\text{Serial execution time}}{\text{Parallel execution time}}$$

Figure 4 shows that the speedup of the GAPS is increased with the increasing number of the search nodes that scale from 1.55 in the case of 2 nodes in order to reach 2.59 in the case of 11 nodes.  The traditional search speedup starts with 1.2 and continues to go up until it reaches 1.9 in when using 5 nodes. However, it starts decreasing to 1.5 when using 11 nodes to perform a search task. This result shows that the GAPS is suitable for large scale dataset size. Furthermore, the GAPS obtained the speed up that is 33% better than the traditional search when two nodes were used. Additionally, the GAPS obtained a better performance of 73% as compared to the traditional search when the number of the nodes is 11.

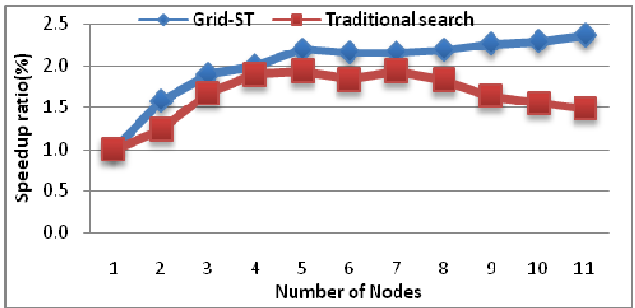

Figure 4: Speedup scales as the increase of size.

*3) Efficiency*

The efficient search technique is a technique that utilizes the resource in a good manner. This experiment is conducted to calculate the GAPS efficiency and to compare its performance with the traditional search. The efficiency is calculated by dividing the speed up by the number of nodes used in the test, which produce an amount of less than 1.0. The perfect and the best efficiency is equal to or very close to 1.0.

Efficiency = Speedup/Number of used nodes

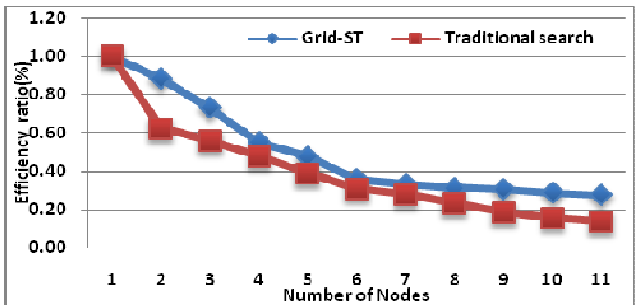

Figure 5: Efficiency scales as the increase of size.

The large dataset experiment reported that the GAPS efficiency started with 0.88 with 2 nodes and decreased to 0.27 with 11 computing nodes. On the other hand, the traditional search efficiency started with 0.62 and decreased until 0.17, as depicted on Figure 5. The result shows that the GAPS has a better efficiency as compared to the traditional search. The GAPS efficiency is 43% better than the traditional search in the case of two nodes, and it is 100% better when using 11 nodes.

## V. CONCLUSION

The paper has highlighted the issues of a massively distributed data and the issues related to the grid-based academic publications search processes. The Grid-based Search Academic Publications technique GAPS was proposed in order to allow the end user to search and access academic publications distributed over a number of organizations. An experiment was performed to measure the performance of the GAPS technique and the suitability of using grid computing as the data sharing infrastructure. The experiment results show that the distribution of the data over the grid and the nodes' capabilities are the key issues that reduce the search performance. Additionally, the GAPS remains the good performance when the system and the publication sources grow for large scale, and it also shows good performance with an increase of publication size.

## REFERENCE

[1] L. Wang, J. Lin, and D. Metzler, "Learning to efficiently rank," in *Proceedings of the 33rd international ACM SIGIR conference on Research and development in information retrieval*, 2010, pp. 138-145.

[2] I. Foster, Y. Zhao, I. Raicu, and S. Y. Lu, "Cloud Computing and Grid Computing 360-Degree Compared," *Gce: 2008 Grid Computing Environments Workshop,* pp. 60-69, 2008.

[3] B. Nicolae, G. Antoniu, L. Bougé, D. Moise, and A. Carpen-Amarie, "BlobSeer: Next-generation data management for large scale infrastructures," *Journal of Parallel and Distributed Computing,* vol. 71, pp. 169-184, 2011.

[4] I. Foster, C. Kesselman, J. Nick, and S. Tuecke, "The physiology of the grid," *Grid computing: making the global infrastructure a reality,* pp. 217–250, 2003.

[5] S. Venugopal, R. Buyya, and K. Ramamohanarao, "A taxonomy of Data Grids for distributed data sharing, management, and processing," *ACM Comput. Surv.,* vol. 38, p. 3, 2006.

[6] J. Smith, P. Watson, A. Gounaris, N. W. Paton, A. A. Fernandes, and R. Sakellariou, "Distributed query processing on the grid," *International Journal of High Performance Computing Applications,* vol. 17, pp. 353-367, 2003.

[7] S. M. Abdulhamid, M. S. A. Latiff, and M. B. Bashir, "On-Demand Grid Provisioning Using Cloud Infrastructures and Related Virtualization Tools: A Survey and Taxonomy," *International Journal of Advanced Studies in Computer Science and Engineering (IJASCSE)*, vol. 3, pp. 49-59. , 2014.

[8] J. Zhang and T. Yang, "Research of Retrieving Model for Digital Library Based on Semantic Grid," in *Information Technology and Applications (IFITA), 2010 International Forum on*, 2010, pp. 431-434.

[9] Z. Jidong and X. Yanzi, "HBUTiGrid: A Knowledge Management Model of Digital Library Based on Semantic Grid," in *Management and Service Science (MASS), 2010 International Conference on*, 2010, pp. 1-4.

[10] L. Yi, "The Application of Semantic Grid in Digital Library Knowledge Management Software Engineering and Knowledge Engineering: Theory and Practice," in *Software Engineering and Knowledge Engineering: Theory and Practice*. vol. 114, Y. Wu, Ed., ed: Springer Berlin / Heidelberg, 2012, pp. 879-886.

[11] N. Nakashole, "A Hybrid Scavenger Grid Approach to Intranet Search," PhD, University of Cape Town, 2009.

[12] A. Chen, L. Di, Y. Bai, Y. Wei, and Y. Liu, "Grid computing enhances standards-compatible geospatial catalogue service," *Computers & Geosciences,* vol. 36, pp. 411-421, 2010.

[13] C.-T. Yang, C.-H. Chen, and M.-F. Yang, "Implementation of a medical image file accessing system in co-allocation data grids," *Future Generation Computer Systems,* vol. 26, pp. 1127-1140, 2010.

[14] C. Town and K. Harrison, "Large-scale grid computing for content-based image retrieval," in *Aslib Proceedings*, 2010, pp. 438-446.